# Quasi-exact expansion of the electron momentum density of the Hooke's atom

Sébastien RAGOT

Laboratoire Structure, Propriété et Modélisation des Solides (CNRS, Unité Mixte de Recherche 85-80). École Centrale Paris, Grande Voie des Vignes, 92295 CHATENAY-MALABRY, FRANCE

## Abstract

An analytic expansion of the exact one-electron momentum density of the Hooke's atom is derived for the case $k = ¼$. Electron correlation is shown to have opposite effects on the momentum density, compared with the Moshinsky's atom, but is qualitatively similar to classical two-electron atomic systems at large momenta.

## 1. Introduction and definitions

The Hooke's atom is a system of two electrons repelling classically while harmonically trapped about a nucleus, for which an exact ground-state wave function can be written in closed-form [1]. This model atom is therefore often used in literature to investigate exact electron distributions [2,3,4,5,6]. So far, the one-electron momentum properties of this model system have however not drawn much attention.

The one-electron momentum density $n(\mathbf{p})$ is related to fundamental functions, which are recalled now [7]. First, the one-electron reduced density-matrix (or 1RDM) derived from a $N$-electron wave function $\psi$ is usually written as [8]:

$$\gamma_1(x;x') = N \int \psi(x, x_2, ..., x_N) \psi^*(x', x_2, ..., x_N) dx_2...dx_N . \tag{1}$$

Integrating eq. (1) over spin variables leads to a "spinless" 1RDM

$$\rho_1(\mathbf{r};\mathbf{r}') = \int [\gamma(x;x')]_{s=s'} ds , \tag{2}$$

or in terms of centre-of-mass ($\mathbf{R} = (\mathbf{r} + \mathbf{r}')/2$) and relative ($\mathbf{s} = \mathbf{r} - \mathbf{r}'$) coordinates

$$\tilde{\rho}_1(\mathbf{R},\mathbf{s}) \equiv \rho_1(\mathbf{r};\mathbf{r}'), \tag{3}$$

The celebrated charge density bypasses information on $\mathbf{s} = \mathbf{r} - \mathbf{r}'$ and reduces to $\rho(\mathbf{R}) = \tilde{\rho}_1(\mathbf{R},0) = \rho_1(\mathbf{R};\mathbf{R})$, while the momentum density is defined [7] as

$$n(\mathbf{p}) = \frac{1}{(2\pi)^3} \int \rho_1(\mathbf{r};\mathbf{r}') e^{i\mathbf{p}\cdot(\mathbf{r}-\mathbf{r}')} d\mathbf{r}d\mathbf{r}' = \frac{1}{(2\pi)^3} \int \tilde{\rho}_1(\mathbf{R},\mathbf{s}) e^{i\mathbf{p}\cdot\mathbf{s}} d\mathbf{R}d\mathbf{s} . \tag{4}$$

Thus, the computation of $n(\mathbf{p})$ requires keeping $\mathbf{s} = \mathbf{r} - \mathbf{r}'$ while averaging information relative to $\mathbf{R}$.

Both one-electron quantities are subjected to the normalization condition

$$\int \tilde{\rho}_1(\mathbf{R},0) d\mathbf{R} = \int \rho(\mathbf{R}) d\mathbf{R} = \int n(\mathbf{p}) d\mathbf{p} = N . \tag{5}$$

The momentum density $n(\mathbf{p})$ allows for computing momentum moments $\langle p^k \rangle = \int p^k n(\mathbf{p}) d\mathbf{p}$, and notably the kinetic energy $T = \frac{1}{2}\langle p^2 \rangle$.

## 2. Exact vs. Hartree-Fock (HF) momentum densities of the Moshinsky's and Hooke's atom

The ground-state singlet HF momentum density of a two-particle system is determined by the unique momentum space orbital $\chi_{HF}$ as $n_{HF}(\mathbf{p}) = 2|\chi_{HF}(\mathbf{p})|^2$ (see for instance ref. [7]). Such an expression reflects the fact that a determinantal 1RDM can be written as a finite sum of orbital products, e.g. only one in the present case. In contrast, the exact 1RDM of a correlated system is known to involve more orbital products, possibly an infinite sum thereof, reflecting the fact that **r** and **r'** are not separable anymore [8].

### 2.1 Moshinsky's atom

This point can be easily understood through a simple example: the Moshinsky's atom, in which two fermions are bounded from a harmonic potential $\frac{1}{2}k(r_1^2 + r_2^2)$ while repelling each other via the Hooke's law $-\frac{1}{2}lr_{12}^2 = -\frac{1}{2}l(\mathbf{r}_1 - \mathbf{r}_2)^2$. Such a system is exactly solvable [9], allowing for the calculation of exact closed-form densities [8], and thus for the comparison of exact and HF density matrices, respectively $\rho_1$ and $\rho_{1,HF}$ given by

$$\rho_1 \propto e^{-(\alpha R^2 + \frac{1}{4}\beta s^2)} = e^{-\frac{1}{4}(\alpha+\beta)(r^2+r'^2) - \frac{1}{2}(\alpha-\beta)\mathbf{r}\cdot\mathbf{r'}},$$

and

$$\rho_{1,HF} \propto e^{-\gamma(R^2 + \frac{1}{4}s^2)} = e^{-\gamma(r^2+r'^2)/2},$$

where $\alpha$, $\beta$, and $\gamma$ are constants defined as

$$\alpha = \frac{2\sqrt{k(k-2l)}}{\left(\sqrt{k} + \sqrt{(k-2l)}\right)},$$

$$\beta = \frac{\sqrt{k(k-2l)} + k - l}{\left(\sqrt{k} + \sqrt{(k-2l)}\right)},$$

and

$$\gamma = \sqrt{(k-l)}.$$

The exact energy $E = \frac{3}{2}\left(\sqrt{k} + \sqrt{(k-2l)}\right)$ requires $k > 2l$ for the particles to remain bounded, contrary to the HF solution $E_{HF} = 3\sqrt{(k-l)}$.

Notice that one-particle densities derived from the 1RDMs above will differ from each other by a simple scale factor. Yet, as $\alpha \neq \beta$ (unless $l = 0$), the exponent in $\rho_1$ is not proportional to $r^2 + r'^2$, so that the exact $\rho_1$ can not be rewritten as a single orbital product, in contrast with $\rho_{1,HF}$. Rather, it may be expanded as an infinite sum of orbital products, e.g. by developing $e^{-\frac{1}{2}(\alpha-\beta)\mathbf{r}\cdot\mathbf{r}'}$ and rearranging expanded terms as products of orbitals.

The HF kinetic energies is 0.612372 atomic units (which are assumed throughout), that is, in error of +3.5%, with respect to the exact kinetic energy (0.591506). Here, virial theorem implies the exact kinetic energy of the Moshinsky's atom in its ground state to be minimal, by virtue of the relation

½ $E = T = U_{ext} + U_{ee}$,

where $U_{ext}$ is the potential energy of electrons in the external harmonic force field while $U_{ee}$ is the electron-electron repulsion energy. Consistently, the difference $n(p) - n_{HF}(p)$ becomes negative at large momenta for the Moshinsky's atom, see figure 1.

### 2.2 Hooke's atom

In the Hooke's atom, the electrons are still bounded from the harmonic potential $\frac{1}{2}k(r_1^2 + r_2^2)$ but now mutually interact via $1/r_{12}$. Using $k = 1/4$ allows the wave function and the charge density to be formulated in closed-form [1,2]. Besides, the wave function can be formulated in momentum space [3]. Taking two-particles momentum variables $\mathbf{P}_{12} = \mathbf{p}_1 + \mathbf{p}_2$ and $\mathbf{p}_{12} = (\mathbf{p}_2 - \mathbf{p}_1)/2$, respectively associated to $\mathbf{R}_{12} = (\mathbf{r}_1 + \mathbf{r}_2)/2$ and $\mathbf{r}_{12} = \mathbf{r}_2 - \mathbf{r}_1$, the momentum wave function can be separated as $\psi = \psi_{P_{12}}(P_{12})\psi_{p_{12}}(p_{12})$, where

$$\psi_{P_{12}}(P_{12}) = \frac{1}{\pi^{3/4}} e^{-P_{12}^{2}/2}, \tag{6}$$

and

$$\psi_{p_{12}}(p_{12}) = \frac{4e^{-2p_{12}^{2}}}{\pi(16+10\pi^{1/2})^{1/2}} \left\{ (2\pi)^{1/2} + 2e^{2p_{12}^{2}}\left(1 - \frac{e^{-2p_{12}^{2}}\sqrt{\pi/2}\left(4p_{12}^{2}-1\right)\mathrm{erfi}\left(\sqrt{2}p_{12}\right)}{2p_{12}}\right) \right\}, \tag{7}$$

where erf*i* is the imaginary error function erf(*ix*)/*i*.

Next, the momentum density can be calculated thanks to the relation

$$n(\mathbf{p}_{1}) = 2\int \left|\psi_{P_{12}}(P_{12})\psi_{p_{12}}(p_{12})\right|^{2} d\mathbf{p}_{2}. \tag{8}$$

However, as a direct integration is not possible due to the second term of $\psi_{p_{12}}$, eq. (7), one may use the following expansion

$$\left(1 - \frac{e^{-2p_{12}^{2}}\sqrt{\pi/2}\left(4p_{12}^{2}-1\right)\mathrm{Erfi}\left(\sqrt{2}p_{12}\right)}{2p_{12}}\right) = 2e^{-2p_{12}^{2}}\sum_{m=0}^{\infty} C_{m} p_{12}^{2m}. \tag{9}$$

where the gaussian in the right-hand term ensures the wave function to be finite integrable and the coefficients $C_m$ are such that expansion coefficients of left and right-hand terms are identical to any order in $p_{12}$. Then, in eq. (8), replacing $\tfrac{1}{2}\mathbf{P}_{12} = \mathbf{p}_{1} + \mathbf{p}_{12}$ and integrating over $d\mathbf{p}_{2} \equiv d\mathbf{p}_{12}$ leads to the following expression for the exact momentum density:

$$n(p) = \frac{K_{1}^{2} e^{-2p^{2}}}{2^{1/2}}$$
$$+ \frac{1}{\pi^{1/2}} K_{1}^{2} K_{2} \sum_{m=0}^{\infty} C_{m} 2^{\frac{3}{2}-3m} \Gamma\left(m+\frac{3}{2}\right) {}_{1}F_{1}\left(m+\frac{3}{2},\frac{3}{2},2p^{2}\right) e^{-4p^{2}} \tag{10}$$
$$+ \frac{1}{\pi^{1/2}} K_{1}^{2} K_{2}^{2} \sum_{m=0}^{\infty}\sum_{n=0}^{\infty} C_{m} C_{n} 2^{\frac{1}{2}-3(m+n)} \Gamma\left(m+n+\frac{3}{2}\right) {}_{1}F_{1}\left(m+n+\frac{3}{2},\frac{3}{2},2p^{2}\right) e^{-4p^{2}}$$

where $K_{1} = \dfrac{4}{\sqrt{8\pi+5\pi^{3/2}}}$, $K_{2} = 2\sqrt{\dfrac{2}{\pi}}$, $\Gamma$ is the Euler gamma function and ${}_1F_1$ is the Kummer confluent hypergeometric function.

Such an expansion of $n(p)$ converges very slowly (see table 1). This is due to the fact that erf$i(z)$ has series about infinity [10] given by $\text{erf}i(z) = \pi^{-1/2} e^{z^2} \left( z^{-1} + \frac{1}{2} z^{-3} + ... \right)$, which asymptotic form is not easily recovered by the expansion of eq. (9).

Accordingly, truncating eq. (9) at any order $m = m_{max}$ leads to some inaccuracy of the gaussian expansion at large values of $p_{12}$. For instance, stopping the expansion at $m_{max} = 4, 16, 64$ and $256$ leads to relative errors of about 1.00, 0.10, 0.01 and 0.001% on the kinetic energy, respectively (the converged value being found to be 0.664417 by numerical integration). This slow convergence of $n(p)$ is somehow a consequence of the inter-electron potential $1/r_{12}$, which reflects in the exact wave function, and in turn makes **r** and **r'** non-separable in the 1RDM.

The HF momentum density can be obtained by expanding the HF orbital over harmonic-oscillator eigenfunctions, following ref. [3] (the five first eigenfunctions were used in the present case). Hence, using the converged value $T = 0.664417$ for the exact kinetic energy, the difference with respect to the HF value amounts to 0.031883 (by numerical integration). The correlation kinetic energy is thus of the same order of magnitude as that of the helium atom (about 0.04204).

The momentum moments of Hooke's atom obtained thanks to the expanded momentum density of eq. (10) are reported in table 2 (accurate to the last digit) together with those of H$^-$ and He atomic systems.

In reference to figures 1 and 2, while the momentum densities look pretty much the same for either the Moshinsky's or Hooke's atom, the differences between exact and HF curves show opposite behaviors. In the Hooke's case (figure 2), $n_{HF}(p)$ overestimates the exact number of electrons of low momenta, consistently with the kinetic energies and other momentum moments reported in table 2. This is further consistent with the implications that the virial theorem has in the present case. Indeed, as discussed in ref. [4,5], the virial theorem implies here

$T = U_{ext} - \frac{1}{2} U_{ee}$.

Thus, since correlation tends to substantially decrease $U_{ee}$, while not affecting too much $U_{ext}$, the exact kinetic energy $T$ must therefore be greater than the HF one. Accordingly, one expects the difference $n(p) - n_{HF}(p)$ to be positive at high momenta (see figure 3 and table 2), owing to the momentum induced by the electrons repelling. Such a phenomenon is in fact rather systematic for coulombic systems of two electrons (see e.g. ref. [11,12]) or even for molecular systems [13,14].

TABLE CAPTIONS & TABLE

Table 1: Values of $m_{max}$, $Cm_{max}$ ($m_{max}$ = truncation value of m in eq. (9), (10), see text), momentum moments $N = \langle p^0 \rangle$ and $T = \frac{1}{2}\langle p^2 \rangle$ of the momentum density, eq. (10), of the Hooke's atom for k = ¼.

Table 2: Comparison between quasi-exact and HF momentum moments of H-, He and the Hooke's atom.

Table 1

| $m_{max}$ | $Cm_{max}$ | $N = \langle p^0 \rangle$ | $T = \frac{1}{2}\langle p^2 \rangle$ |
|---|---|---|---|
| 0 | 1. | 2.833029 | 1.06239 |
| 1 | -6.666667 $10^{-1}$ | 2.073745 | 0.696685 |
| 2 | -1.333333 $10^{-1}$ | 2.00832 | 0.661771 |
| 3 | -3.809524 $10^{-2}$ | 1.997905 | 0.656692 |
| 4 | -1.058201 $10^{-2}$ | 1.996832 | 0.657177 |
| 8 | -2.489885 $10^{-5}$ | 1.999073 | 0.661649 |
| 16 | -3.061856 $10^{-12}$ | 1.999873 | 0.663651 |
| 32 | -3.985972 $10^{-30}$ | 1.999981 | 0.664179 |
| 64 | -8.873795 $10^{-75}$ | 1.999997 | 0.664339 |
| 128 | -1.346499 $10^{-182}$ | 1.999999 | 0.6643909 |
| 256 | -5.149269 $10^{-436}$ | 2. | 0.6644084 |
| ⋮ |  | ⋮ | ⋮ |
| ∞ |  | 2 | 0.664417 |

Table 2

| | H- [15,11] | | | He [15,16] | | | Hooke's atom (this work) | | |
|---|---|---|---|---|---|---|---|---|---|
| | "exact" | HF | | "exact" | HF | | "exact" | HF | |
| $\langle p^{-2} \rangle$ | 42,900 | 34,571 | 19,4 % | 4,099 | 4,092 | 0,2 % | 9,04965 | 9,29640 | -2,7 % |
| $\langle p^{-1} \rangle$ | 6,446 | 5,999 | 6,9 % | 2,139 | 2,141 | -0,1 % | 3,39636 | 3,44843 | -1,5 % |
| $\langle p^{1} \rangle$ | 1,115 | 1,098 | 1,5 % | 2,815 | 2,799 | 0,6 % | 1,49986 | 1,46947 | 2,0 % |
| $\langle p^{2} \rangle$ | 1,055 | 0,976 | 7,5 % | 5,807 | 5,723 | 1,4 % | 1,3288 | 1,26507 | 4,8 % |
| $\langle p^{3} \rangle$ | 1,658 | 1,458 | 12,1 % | 18,406 | 17,990 | 2,3 % | 1,343 | 1,22546 | 8,7 % |

FIGURE CAPTIONS & FIGURES

Fig. 1: radial momentum densities of the Moshinsky atom ($k = 1/4$, $l = 1/12$): exact (full line), HF (large dashed), exact - HF (small dashed).

Fig. 2: radial momentum densities for the Hooke's atom ($k = 1/4$). Upper fig.: quasi-exact (full line, from eq. (10), stopping the expansion at $m_{max} = 64$), HF (large dashed, see ref. [3]). Lower fig.: quasi-exact - HF.

Fig. 1

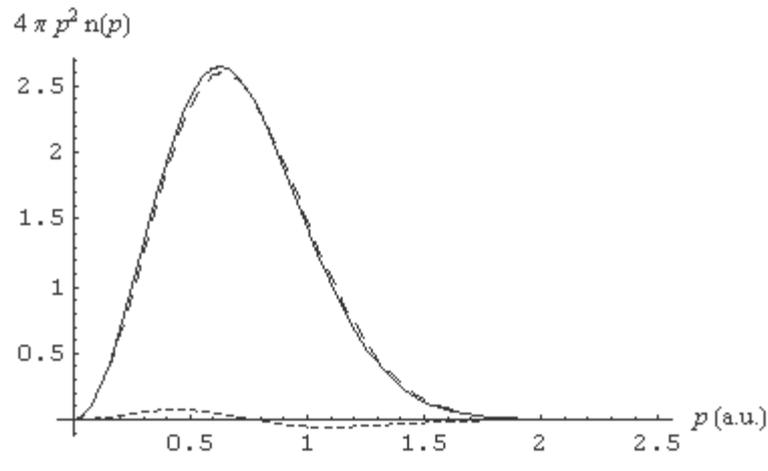

Fig. 2

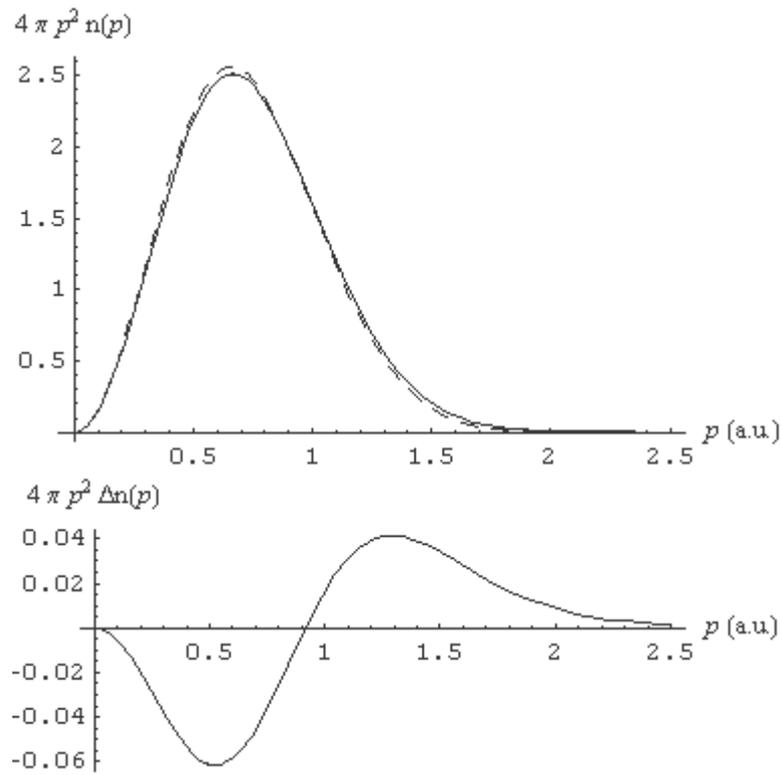